\documentclass[aps,prb,twocolumn,groupedaddress,showpacs]{revtex4}
\usepackage{graphicx}% Include figure files
\usepackage{dcolumn}% Align table columns on decimal point
\usepackage{bm}% bold math
\begin{document}

%\preprint{}

\title{Complex domain-wall dynamics in compressively strained Ga$_{1-x}$Mn$_{x}$As epilayers\\}

\author{Liza Herrera Diez}
 \affiliation{%
 Max-Planck-Institut f\"ur
Festk\"orperforschung, Heisenbergstrasse 1, 70569, Stuttgart, Germany}% Lines break automatically or can be forced with \\

\author{Reinhard. K. Kremer}%
\affiliation{%
Max-Planck-Institut f\"ur
Festk\"orperforschung, Heisenbergstrasse 1, 70569, Stuttgart, Germany}% Lines break automatically or can be forced with \\

\author{Axel Enders}%
\altaffiliation[Present address: ] {Department of Physics and
Astronomy, University of Nebraska, Lincoln, NE 68588, USA.}
\affiliation{%
Max-Planck-Institut f\"ur
Festk\"orperforschung, Heisenbergstrasse 1, 70569, Stuttgart, Germany}% Lines break automatically or can be forced with \\

\author{Matthias R\"ossle}%
\altaffiliation[Present address: ] {Department of Physics,
University of Fribourg, Chemin du Mus\'ee 3, CH-1700 Fribourg,
Switzerland.}
\affiliation{%
Max-Planck-Institut f\"ur
Festk\"orperforschung, Heisenbergstrasse 1, 70569, Stuttgart, Germany}% Lines break automatically or can be forced with \\

\author{Erhan Arac}%
\altaffiliation[Present address: ] {Forschungszentrum Karlsruhe,
Institut f\"ur Festk\"orperphysik, 76021 Karlsruhe, Germany.}
\affiliation{%
Max-Planck-Institut f\"ur
Festk\"orperforschung, Heisenbergstrasse 1, 70569, Stuttgart, Germany}% Lines break automatically or can be forced with \\

\author{Jan Honolka}%
\email{j.honolka@fkf.mpg.de}
\affiliation{%
Max-Planck-Institut f\"ur
Festk\"orperforschung, Heisenbergstrasse 1, 70569, Stuttgart, Germany}% Lines break automatically or can be forced with \\

\author{Klaus Kern}%
\affiliation{%
Max-Planck-Institut f\"ur
Festk\"orperforschung, Heisenbergstrasse 1, 70569, Stuttgart, Germany}% Lines break automatically or can be forced with \\

\author{Ernesto Placidi}%
\affiliation{%
Dipartimento di Fisica, Universit\`a di Roma "Tor Vergata", and
CNR-INFM, Via della Ricerca Scientifica 1, I-00133
Roma, Italy}% Lines break automatically or can be forced with \\

\author{Fabrizio Arciprete}%
\affiliation{%
Dipartimento di Fisica, Universit\`a di Roma "Tor Vergata", and
CNR-INFM, Via della Ricerca Scientifica 1, I-00133
Roma, Italy}% Lines break automatically or can be forced with \\

\author{}
%\thanks{}
\affiliation{}


\date{\today}

\begin{abstract}
The domain wall induced reversal dynamics in compressively strained
Ga$_{1-x}$Mn$_{x}$As was studied employing the magneto-optical Kerr
effect and Kerr microscopy. Due to the influence of an uniaxial part
in the in-plane magnetic anisotropy 90$^{\circ}\pm \delta$ domain
walls with considerably different dynamic behavior are observed.
While the 90$^{\circ} +\delta$ reversal is identified to be
propagation dominated with a small number of domains, the case of
90$^{\circ}-\delta$ reversal involves a larger number of nucleation
centers. The domain wall nucleation/propagation energies $\epsilon$
for both transitions are estimated using model calculations from
which we conclude that single domain devices can be achievable using
the 90$^{\circ}+\delta$ mode.
\end{abstract}

\pacs{75.50.Pp, 75.60.Ch, 75.60.Jk}
\keywords{GaMnAs, domain wall dynamics, magnetic anisotropy}

\maketitle The discovery of the ferromagnetic semiconductor
Ga$_{1-x}$Mn$_{x}$As and the possible implementation into spintronic
devices triggered great interest in understanding its fundamental
properties~\cite{ohno}. The linkage between carrier density and
magnetic properties in this hole mediated ferromagnet allows tuning
of its magnetic properties such as the Curie temperature ($T_c$)
upon changing the carrier concentration~\cite{dietl,chiba}. In
addition, magnetic domain wall (DW) logic operations may be
implemented~\cite{allwood} including magneto-resistive read-outs.
However, any application in this direction requires a full control
over magnetic reversal dynamics, which in most cases happens via the
nucleation and propagation of domain walls.\\
A good understanding of the magnetic anisotropy landscape is also
required not only for the design of magneto-resistive devices but
also because magnetic anisotropy can manifest in the domain wall
dynamics. The magnitude of the magnetic anisotropy is related to
important parameters such as the domain wall energy and
width~\cite{book} which can determine a process to be propagation or
nucleation dominated. This can be very well visualized in the effect
of a nonuniform anisotropy distribution in simulated reverse domain
patterns~\cite{montecarlo}.\\
So far, domain wall studies by means of Kerr microscopy (KM) in
Ga$_{1-x}$Mn$_{x}$As have been mostly performed in films with
tensile strain where the magnetization pointed perpendicular to the
plane~\cite{yamanouchi}. Ga$_{1-x}$Mn$_{x}$As with in-plane
magnetization has been extensively studied using magneto-transport
measurements~\cite{pappert,tang}, however, this techniques does not
provide spatially resolved information about DW nucleation and
propagation processes. In this study we present the direct
observation of DW motion in compressively strained
Ga$_{1-x}$Mn$_{x}$As by KM and the dependence of the DW dynamics on
the direction of the applied magnetic field. While an earlier
magneto-optical study in the literature~\cite{welp} did not address
possible anisotropies in the DW dynamics our KM results reveal a
distinct anisotropy in the DW dynamics dependent on the direction of
the applied magnetic field with respect to the crystal axes. From
the analysis of angle resolved magneto-optical Kerr effect(MOKE)
measurements we attribute this anisotropy to the existence of two
different types of DWs. All measurements were done on Hall Bar
devices of 150$\mu$m width, patterned in [1$\bar{1}$0]
and [110] directions using photolithography. \\
The Ga$_{1-x}$Mn$_{x}$As sample was grown in a RIBER 32 molecular
beam epitaxy (MBE) system equipped with a reflection high-energy
electron diffraction (RHEED) setup for \emph{in situ} monitoring of
the growth. Prior to Ga$_{1-x}$Mn$_{x}$As deposition, a GaAs buffer
layer of approximately 400 nm was grown on a Si-doped GaAs(001)
substrate (n $\approx$ $10^{18}$ cm$^{-3}$), in As$_{4}$ overflow at
$\approx$ 590 $^{\circ}$C, and at a rate of 0.8 $\mu$m/h. After 10
minutes post-growth annealing under As$_{4}$ flux, the temperature
was lowered to 270 $^{\circ}$C for Ga$_{1-x}$Mn$_{x}$As deposition.
Using an As$_{4}$:Ga flux ratio of 50 a 170 nm Ga$_{1-x}$Mn$_{x}$As
layer was grown at a rate of 0.33 ML/sec. During
Ga$_{1-x}$Mn$_{x}$As growth a clear two-dimensional (1$\times$2)
RHEED pattern was observed with no indication of MnAs precipitates
at the surface (no spotty RHEED pattern). Ga, Mn and As$_{4}$ fluxes
were calibrated by an ion gauge placed in the substrate position
(beam equivalent pressure - BEP). A nominal Mn concentration of
$\emph{x}$ = (2.3 $\pm$ 0.1) $\%$ was estimated on the basis of flux
(BEP) ratios of As$_{4}$, Ga, and Mn. The high quality of the grown
films has also been verified by measuring the X-ray diffraction
pattern of the film. Typical diffraction profiles for
Ga$_{1-x}$Mn$_{x}$As/GaAs structures have been found, which contain
two distinct peaks around the (004) Bragg reflex corresponding to
the GaAs and Ga$_{1-x}$Mn$_{x}$As layers~\cite{sadowsky}.\\
Longitudinal MOKE measurements have been done at a temperature of $T
\sim 3$K changing the direction of the in-plane applied field with
respect to the crystal axes in order to map the coercivities and
thus the magnetic anisotropy. Along the [110] and [1$\bar{1}$0]
directions we observe only one switching field within the available
field range, while in other directions as in the case of the [100]
direction (Fig.1 (a)), two transitions and an intermediate plateau
were found. The observed dependence of the switching fields on the
direction of the applied field are consistent with the results
already found by other authors~\cite{tang} in Ga$_{1-x}$Mn$_{x}$As
with similar Mn concentration. In addition, the magnetization value
as a function of temperature presented in Fig.1 (b) was measured
using a SQUID magnetometer. The Curie temperature and saturation
magnetization values
determined from this measurement are $T_{\text{C}} =(48\pm 2)$~K and $\emph{M} = (9.4\pm 0.1)$~emu/cm$^{3}$, respectively.\\
\begin{figure}
  \includegraphics[width=9cm]{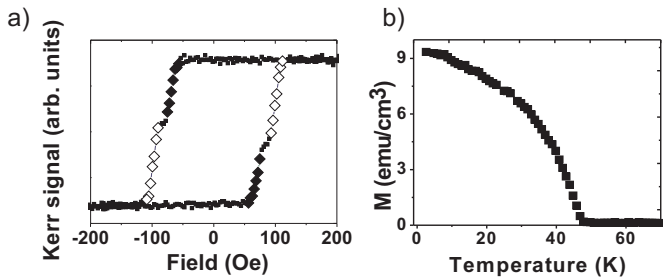}\\
  \vspace{-10pt}
  \caption{(a) Two step hysteresis loop measured with the field applied
along [100] at $T=3$ K. The first and second switching events are
indicated by filled and open symbols, respectively. (b)
Magnetization as a function of temperature measured at $H=1$ T with
SQUID in zero field cooling.}\label{xx}
\end{figure}
Prior to every single KM measurement the sample was saturated at a
magnetic field of 1000 Oe. The transition was then triggered by
applying a constant field of opposite polarity corresponding to the
respective switching field. Fig.2 shows Kerr images\cite{horcas} of
the domain wall transitions mediating the magnetization reversal for
the field applied along the [1$\bar{1}$0] (a) and [110] (b)
directions, respectively. They correspond to hysteresis curves with
single switching events displayed in Fig.2 (c). The two Kerr images
left and right in (a) and (b) were taken at consecutive times. A
clear asymmetry in the nucleation behaviour is observed. For the
field applied in the [1$\bar{1}$0] direction (Fig.2 (a)) the reverse
domains nucleate in large numbers. In contrast, when the field is
applied in the [110] direction (b) the transition is dominated by
the
propagation of a few DWs nucleated at the contact pads.\\
\begin{figure}
  \includegraphics[width=7cm]{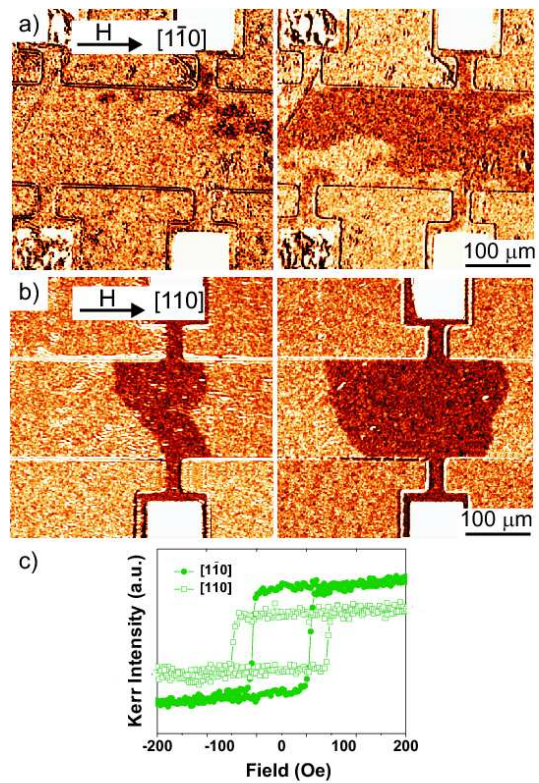}\\
  \vspace{-10pt}
  \caption{ DW nucleation and propagation for the field
applied in the [1$\bar{1}$0] (a) and [110] (b) directions at two
consecutive times left and right. The respective hysteresis loops
showing single switching events are displayed in (c). A larger
number of domains is
found when applying the field along the [1$\bar{1}$0] direction.}\label{xx}
\end{figure}

In the following we will present evidence supporting the existence
of two different kind of DWs which seem to be the cause for the
asymmetry in the reversal behaviour. These two DW types are given by
the interplay of the uniaxial and biaxial magnetic anisotropy in
compressively strained Ga$_{1-x}$Mn$_{x}$As
epilayers~\cite{Hrabovsky,wang}. The biaxial component is a
spin-orbit coupling effect well described by the theory of hole
mediated ferromagnetism. The origin of the uniaxial anisotropy has
been related to a small trigonal lattice distortion but the
mechanism leading to this symmetry breaking still remains
unclear~\cite{jungwirth,Sawicki}.\\
Assuming a fourfold crystalline anisotropy plus an uniaxial
contribution the energy of a in-plane single domain state can be
described by:
\begin{equation}
\label{energy}
E=\frac{K_{c}}{4}\sin^{2}(2\varphi) + K_{u}\sin^{2}(\varphi-135^{\circ}) - M H\cos(\varphi-\varphi_{H})
\end{equation}
where $K_{c}$ and $K_{u}$ are the biaxial and uniaxial anisotropy
constants, $\emph{M}$ is the magnetization, $\emph{H}$ the magnetic
field, and $ \varphi$ and $ \varphi_{H}$ are the angles of
$\emph{M}$ and $\emph{H}$ with the [100] direction. The presence of
an uniaxial easy axis along the [110] direction shifts the position
of the energy minima from the [100] and [010] directions, which are
the easy axes in the case of a pure biaxial anisotropy, towards the
[110] axis. This shift in angle $\delta/2$ is determined by the
ratio between the biaxial and uniaxial anisotropies in the following
manner~\cite{pappert,daboo}:
\begin{equation}
\label{energy}
\frac{\delta}{2}=\frac{1}{2}\arcsin(\frac{K_{u}}{K_{c}})
\end{equation}
The value $\delta/2$ can be geometrically derived using the polar
plot of the switching fields in the inset of Fig.3 . $\delta/2$ is
determined by the angle difference between the corners of the
rectangle defined by the first switching fields (green circles)
close to the [100] and [010] directions~\cite{molenkamp}. These
corner points are also the directions where first and second
switching fields coincide. We obtain $\delta/2 =($15$\pm 2)
^{\circ}$, indicating that the global easy axes are located ($60 \pm
2) ^{\circ}$ from the [1$\bar{1}$0] and ($30 \pm 2) ^{\circ}$ from
the [110] direction, respectively.\\
\begin{figure}
  \vspace{-10pt}
  \includegraphics[width=8cm]{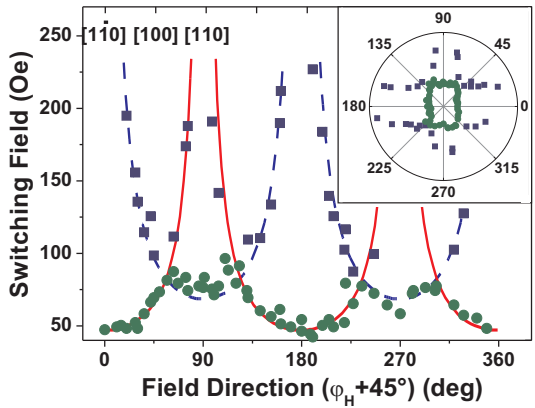}\\
  \vspace{-20pt}
  \caption{ Fits of the experimental results of first
(green circles) and second (blue squares) switching field vs. field
angle. The blue dashed and red solid lines represent fits to
Eq.(~\ref{switching_field}) considering 90$^{\circ}$+$\delta$ and
90$^{\circ}$-$\delta$ DWs, respectively. The value of $\delta/2$ was
obtained from the data points as indicated. The data points form two
sets of parallel lines when plotted in polar coordinates (inset).}\label{xx}
\end{figure}
In order to fit the experimental values of the switching field vs.
the direction of applied field presented in Fig.3, the DW
nucleation/propagation energy $\epsilon$ was equated with the gain
in Zeeman energy during the reversal between the initial
($\textbf{M$_{0}$}$) and final ($\textbf{M$_{1}$}$) state of the
magnetization in a constant field,
$\epsilon$=$\textbf{H}\cdot$($\textbf{M$_{1}$}$-$\textbf{M$_{0}$}$).
It has been previously shown~\cite{pappert} that in this case the
angle change in the magnetization direction during the reversal can
be either 90$^{\circ}$+$\delta$ or 90$^{\circ}$-$\delta$ reflecting
the shift in the easy axes directions by the angle $\delta/2$.
Consequently two expression accounting for 90$^{\circ}$+$\delta$ and
90$^{\circ}$-$\delta$ transitions have to be considered with
corresponding $\epsilon_{90+\delta}$ and $\epsilon_{90-\delta}$.
These two expressions for the DW transition ($\epsilon_{90-\delta}$=
$\textbf{H}\cdot$($\textbf{M$_{1}$}$-$\textbf{M$_{0}$}$),
$\epsilon_{90+\delta}$=
$\textbf{H}\cdot$($\textbf{M'$_{1}$}$-$\textbf{M'$_{0}$}$)) yield a
dependence of the switching field on the angle of the applied field
for 90$^{\circ} \pm \delta$ DWs:
\begin{equation}
\label{switching_field}
H_{90\pm\delta}= \frac{\epsilon_{90\pm\delta}}{\emph{M} \sqrt{2}
\cos(45\mp\frac{\delta}{2}) (\sin (\varphi_{H})\mp
\cos(\varphi_{H}))}
\end{equation}
Using $\delta/2 =($15$\pm 2) ^{\circ}$ and $\emph{M}$ =
(9.4$\pm$0.1) emu/cm$^{3}$ we are able to fit the data in Fig.3
according to Eq.(3) with $\epsilon_{90+\delta}$ and
$\epsilon_{90-\delta}$ as fitting parameters. The two functions
shown in Fig.3 very well reproduce the two branches of the switching
field observed in the MOKE measurement. The solid line represents
the 90$^{\circ}$-$\delta$ (60$^{\circ}$ DW) transition with the
corresponding value for the nucleation/propagation energy of
$\epsilon_{90-\delta}$= 460 J/cm$^{3}$. The dashed line models the
90$^{\circ}$+$\delta$ (120$^{\circ}$ DW) reversal process with
$\epsilon_{90+\delta}$= 1173 J/cm$^{3}$. Taking this into account,
the switching field obtained with the field oriented along the
[1$\bar{1}$0] direction would correspond to a 60$^{\circ}$ DW since
the experimental value for the switching field lays on the solid fit
curve. Similarly, for the field applied in the [110] axis, the
switching field is found on the dashed fit curve and consequently
corresponds to a 120$^{\circ}$ DW. As discussed earlier these two
types of DWs seem to show
a very different nucleation/propagation behavior.\\
According to Fig.3 the second transition can not be triggered within
the available magnetic field range when the fields are applied
either along the [1$\bar{1}$0] or the [110] directions. However, to
corroborate the observations for the field applied along these two
directions, the nucleation has been studied for fields applied in
the [100] direction. In this case the longitudinal axis of the Hall
Bar was fixed at an angle of 45$^{\circ}$ with respect to the
direction of the field. According to the plot of switching field vs.
field angle in Fig.3, in the [100] direction ($\varphi_{H} =
0^{\circ}$) both of the two transitions can be observed, which add
up to a full 180$^{\circ}$ reorientation. The two switching events
are clearly visible in the MOKE signal in Fig.1 (a). According to
the fit in Fig.3, the first switching field along the [100] axis
corresponds to a $60^{\circ}$ DW transition and the second to a
$120^{\circ}$ DW. In analogy with the previously shown dynamics a
larger number of nucleation centers are expected for the first
switching field (lying on the solid line of the fitting curve) and
only few domains for the second switching field (lying on the dashed
line of the fitting curve) suggesting a DW propagation dominated
reversal. The results shown in Fig.4 (a) and (b), (first and second
transition, respectively) confirm this prognosis and therefore
support the notion of the presence of two species of DWs with
different dynamics. Fig.4 (a) corresponds to the first jump in the
Kerr signal found after saturation as indicated with filled symbols
in Fig.1 (a). Fig.4 (b) shows the
subsequent single domain wall transition indicated by the open symbols, respectively.\\
\begin{figure}
  \includegraphics[width=7cm]{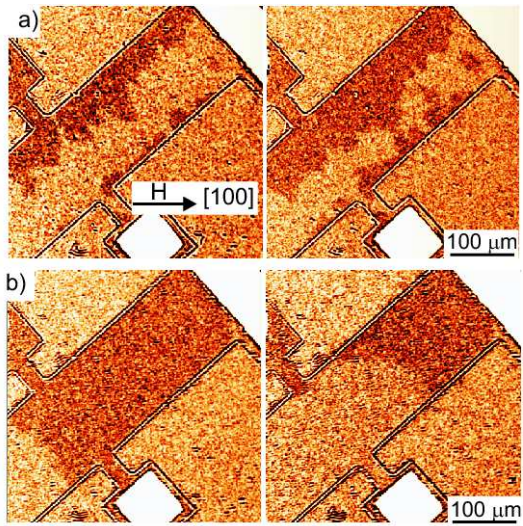}\\
  \caption{ Kerr images of DWs involved in the first (a)
and second (b) switching event for the field applied along the [100]
direction at two consecutive times left and right. In the MOKE
hysteresis loop an intermediate plateau appears between the first
and the second switching event (see Fig.1 (a)).}\label{xx}
\end{figure}
From the fitting we extracted a ratio of $\epsilon$ for the two
types of DWs of $(\epsilon_{90+\delta}/\epsilon_{90-\delta}) = 2.5$.
The experimental results thus indicate a considerably lower
nucleation/propagation energy for the 90$^{\circ}-\delta$ transition
with respect to 90$^{\circ}+\delta$. At the same time KM proves that
low $\epsilon$ values are correlated with a larger number of
nucleation sites. We therefore conclude that the dynamics shifts
from a propagation dominated regime to a partly nucleation dominated
one for high and low $\epsilon$, respectively. In the former case
the observed low number of domains is most likely determined by a
few isolated
defects within the Hall devices serving as nucleation centers. \\
Finally, we shortly want to discuss the special quantity $\epsilon/M$
given by the value of the switching field at exactly the crossing point
of the fitting lines ($\varphi_{H} = 15^{\circ}$) where only one switching
event occurs. As discussed earlier, this point defines the global easy axes
direction of the system. The quantity $\epsilon/M$ is of interest for a
comparison of the material used here with materials grown in other laboratories.
Gould {\it et al.} give a list of values with $\epsilon/M$ ranging between = 7.1
and 18~mT for Ga$_{1-x}$Mn$_{x}$As materials grown in various well-known MBE
facilities~\cite{molenkamp}. Our value $\epsilon/M$ = 8.7 mT falls well into
that range confirming the comparability of our material to other high-quality
materials in the literature.\\
In conclusion, extensive MOKE and Kerr-microscopy studies of the
nucleation of domains and the propagation behavior of DWs in
Ga$_{1-x}$Mn$_{x}$As Hall bar devices revealed substantially
different dynamics for two observed species of DWs. They correspond
to 90$^{\circ} \pm \delta$ DWs and they originate from the uniaxial
part of the magnetic anisotropy. While the 90$^{\circ}+\delta$
reversal is found to be propagation dominated with a small number of
domains, the 90$^{\circ}-\delta$ case involves a larger number of
nucleation centers. The measured coercivities for both reversals can
be very well fitted by a model which includes the uniaxial
anisotropy contribution. From the fits a considerably lower
nucleation/propagation energy $\epsilon$ for the 90$^{\circ}-\delta$
transition with respect to the 90$^{\circ}+\delta$ transition is
derived, which suggests an inverse correlation of $\epsilon$ with
the number of nucleated domains. In the case of 90$^{\circ}+\delta$
the observed low number of domains is determined by few defects
within the Hall devices serving as nucleation centers, which opens
the possibility to design single domain devices e.g. for
magneto-logic elements.
\begin{acknowledgments}
We wish to thank Dr. Dinnebier for X-ray diffraction and P. Kopold
for TEM measurements.
\end{acknowledgments}

\end{document}